\documentstyle[prl,aps,epsf,floats,preprint]{revtex}
\newcommand\LQCD{{\Lambda_{\rm QCD}}}
\newcommand\sla[1]{#1\hskip-0.5em \slash} 
\newcommand\Sla[1]{#1\hskip-0.7em \slash} 
\newcommand\vev[1]{\langle{#1}\rangle}
\newcommand\hq{{\hat q}}
\newcommand\hm{{\hat m}}

\newcommand\ie{{\it i.e.}}
\newcommand\beq{\begin{equation}}
\newcommand\eeq{\end{equation}}
\newcommand\hbv{{h^{(b)}_{v^{\phantom{\prime}}}}}
\newcommand\barhbv{{\bar h^{(b)}_{v^{\phantom{\prime}}}}}
\newcommand\muz{{\mu_0^{\phantom{\dagger}}}}
\begin{document}

\title{Short Distance Analysis of  $\bar B\to D^{(\ast)0} e^+e^-$ and
$\bar B\to J/\psi e^+e^-$}
\author{David H. Evans\footnote{daevans@physics.ucsd.edu}, Benjam\'{\i}n
Grinstein\footnote{bgrinstein@ucsd.edu} and Detlef
R. Nolte\footnote{dnolte@ucsd.edu} \\[4pt]} \address{\tighten{\it
Department of Physics,\\ University of California at San Diego, La
Jolla, CA 92093 USA}\\[4pt] UCSD/PTH 99--09}

\maketitle

\begin{abstract}
Over a large fraction of phase space a
combination of an operator product and heavy quark expansions
effectively turn the decay $\bar B\to D^{(\ast)0} e^+e^-$ into a
``short distance'' process, \ie, one in which the weak and
electromagnetic interactions occur through single local
operators. These
processes have an underlying W-exchange
quark diagram topology and are therefore Cabibbo allowed but 
suppressed by combinatoric
factors and short distance QCD corrections. Our technique allows a
clearer exploration of these effects. For the decay $\bar B_{d,s}\to
J/\psi(\eta_c) e^+e^-$ one must use a non-relativistic (NRQCD)
expansion, in addition to an operator product expansion and a heavy
quark effective theory expansion.  We estimate the decay rates for
$\bar B_{d,s}\to J/\psi e^+e^-$, $\bar B_{d,s}\to \eta_c e^+e^-$,
$\bar B_{d,s}\to D^{*0} e^+e^-$ and $\bar B_{d,s}\to D^{0} e^+e^-$.

\end{abstract}


\vspace{0.2in}

\newpage

\section{Introduction}
In a recent paper\cite{evans-99-2} we considered the collection of
decays $B^+\to D_{s,d}^{(*)+}e^+e^-$. The decay rate for these is
proportional to $|V_{ub}|^2$. We found that over a large kinematic
domain one can reliably estimate the rate (in terms of
$|V_{ub}|^2$). The process is first order weak and first order
electromagnetic, and, therefore, the amplitude involves long
distance physics. The central observation of \cite{evans-99-2} is that
over a large kinematic domain the interaction is local on the scale of
strong dynamics. The amplitude can, therefore, be approximated by the
matrix elements of local operators, which can be estimated in a
variety of ways and should eventually be determined in numerical
simulations of QCD on the lattice. 
The branching fraction for $B^+\to D_{s}^{*+}e^+e^-$, restricted to
invariant mass of the $e^+e^-$ pair in excess of $1.0$~GeV, was
estimated to be $1.9\times10^{-9}$. This is too small to be measured
in $e^+e^-$ B-factories, but could be observable at high luminosity
high energy hadronic colliders. 

\begin{figure}
\centerline{
\epsfysize 2.0in
\epsfbox{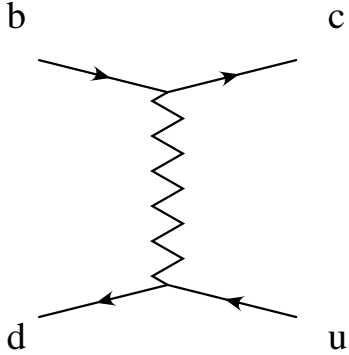}}
\vskip0.5cm
\caption{W-exchange quark topology diagram  underlying the
transition $\bar B_{d,s}\to D^{(*)0} e^+e^-$. Emission of a $ e^+e^-$
pair from any line is understood.}
\label{fig:fig0}
\end{figure}

In this paper we consider the decays $\bar B_{s,d}\to J/\psi e^+e^-$, 
$\bar B_{s,d}\to \eta_c e^+e^-$, 
$\bar B_{s,d}\to D^{*0} e^+e^-$ and $\bar B_{s,d}\to D^{0}
e^+e^-$. These proceed via
W-exchange topologies, as shown in Fig.~\ref{fig:fig0}. In addition,
$\bar B_{d,s}\to J/\psi e^+e^-$ and $\bar B_{d,s}\to \eta_c e^+e^-$ have
small contributions from penguins, which we neglect. 
The goal of the paper is to show how the methods introduced in 
paper\cite{evans-99-2} can be applied to the processes considered here. 
The kinematics of $\bar
B_{d,s}\to D^{(*)0} e^+e^-$ is similar to that of $B^+\to
D_{d,s}^{(*)+}e^+e^-$ so one expects the methods to apply readily. In
fact, the only dynamical difference is that in $\bar B_{d,s}\to
D^{(*)0} e^+e^-$ the heavy $b$ quark decays to a heavy $c$-quark, whereas in
$B^+\to D_{s,d}^{(*)+}e^+e^-$ it is a heavy $b$-anti-quark that decays into
a heavy $c$-quark. The case $\bar B_{d,s}\to J/\psi(\eta_c) e^+e^-$ is
clearly different: both quark and anti-quark in the final state are
heavy and they are moving together in a  bound charmonium state. As we
will see the expansion that arises naturally corresponds to NRQCD, the
non-relativistic limit of heavy quarks bound by QCD  into quarkonia.

The processes under consideration here have advantages compared to
$B^+\to D_{s,d}^{(*)+}e^+e^-$.  These processes are not suppressed by
the small CKM element $|V_{ub}|^2$. One might hope that the decay rate
is, therefore, substantially higher.  However, the
enhancement of the rate due to bigger CKM elements is partially
cancelled by small Wilson coefficients. Therefore, all these processes
have small branching fractions. While none are observable at
B-factories, some are 
observable at future hadronic collider experiments like LHC-B and
BTeV.

These processes are first order weak and first order electromagnetic,
and, therefore, the amplitude involves long distance physics. We will
show that over a large kinematic domain the interaction is
approximated by a set of matrix elements of local operators. All
these matrix elements should eventually be determined by lattice
calculations. For the processes considered in this paper, the number of
independent matrix elements is reduced by the use of
rotational, heavy quark spin and chiral symmetries.

This paper is organized as follows. In Sec.~\ref{sec:method} we review
the methods of Ref.~\cite{evans-99-2} that lead to an expansion in
local operators. The review is done in terms of the graphs relevant to
$\bar B\to D^0 e^+e^-$, which is one of the processes of interest
here. In Sec.~\ref{sec:spinsym} we present a novel analysis that shows
that the matrix elements of
the operators in the expansion are all related by a combination of
heavy-spin, rotational and chiral symmetries. We then proceed to find
the short distance QCD corrections to our operator expansion in
Sec.~\ref{sec:QCD1}. In Sec.~\ref{sec:results1} we give expressions for 
the differential decay rates in terms of matrix elements of local
operators. 
These should be considered our main results. 
To get some numerical estimates of the decay rates
we crudely approximate the local  matrix elements. The
material in Secs.~\ref{sec:method}--\ref{sec:results1} deals with the
decays $\bar B_q\to D^{*0} e^+e^-$ and $\bar B_q\to D^{0} e^+e^-$, and
we repeat the steps  applied to the processes $B_q\to\eta_c
e^+e^-$ and $B_q\to J/\psi e^+e^-$ in Sec.~\ref{sec:onium}.
Our results are  summarized in Sec.~\ref{sec:conclusions}.

\section{Operator Expansion}
\label{sec:method}
In this section we review the method introduced in
\cite{evans-99-2}. However, we will present the method as applied to
the process $\bar B_{d}\to D^{(*)0} e^+e^-$. Therefore we will at once
review the method and perform the necessary calculation for one of the
cases of interest.

\begin{figure}
\centerline{
\epsfysize 2.0in
\epsfbox{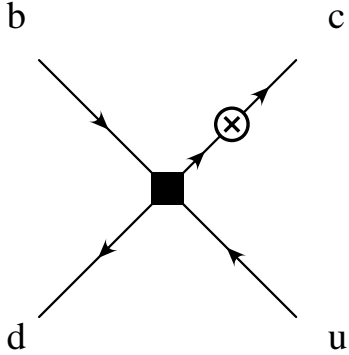}}
\vskip0.5cm
\caption{Feynman diagram representing a contribution to the Green
function. The filled square represents the four quark operator ${\cal
O}$ and the cross represents the electromagnetic current $j^\mu_{{\rm
em}}$, cf. Eq.~(\ref{T-prod}), which here couples to the $c$-quark. }
\label{fig:fig1}
\end{figure}

The effective
Hamiltonian for the weak transition in $\bar B_{d}\to
D^{(*)0} e^+e^-$, is 
\beq
\label{eq:Heff}
{\cal H}'_{\rm eff}= \frac{4G_F}{\sqrt2}\,V^{\phantom{*}}_{ud}V^*_{cb}\left(
c(\mu/M_W){\cal O}+c_8(\mu/M_W){\cal O}_8\right),
\eeq
where 
\beq
\label{eq:Odefd}
{\cal O}=\bar d\gamma^\nu P_- b \;\;\bar c\gamma_\nu P_- u
\eeq
and
\beq
{\cal O}_8=\bar d\gamma^\nu P_-T^a b\;\; 
\bar c\gamma_\nu P_- T^a u,
\eeq
$P_\pm\equiv(1\pm\gamma_5)/2$ and $T^a$ are the generators of color
gauge symmetry. This is a useful basis of operators for our purposes
since  the hadronic matrix element of the ``octet'' operator ${\cal
O}_8$ is suppressed.  The dependence on the renormalization point $\mu$ of
the short distance coefficients $c$ and $c_8$ cancels the
$\mu$-dependence of operators, so matrix elements of the
effective Hamiltonian are $\mu$-independent. 

The amplitude for $\bar B_{d}\to D^{(*)0} e^+e^-$, to leading order in
weak and electromagnetic interactions and to all orders in the strong
interactions involves the following non-local matrix element:
\beq
\label{T-prod}
\langle D^{*+}| \int d^4x\;e^{iq\cdot x}
\; T(j^\mu_{\rm em}(x){\cal O}(0)) |B^+\rangle.
\eeq
Here $q$ denotes the momentum of the $e^+e^-$ pair, $j^\mu_{\rm em}$
is the electromagnetic current operator and the operator ${\cal O}$,
defined in Eq.~(\ref{eq:Odefd}), is the long distance approximation to
the $W$-exchange graph. The full amplitude will of course also involve
a similar non-local matrix element but with the ``singlet'' operator
${\cal O}$ replaced by the octet operator ${\cal O}_8$. For now we
concentrate on the singlet operator. None of the arguments given in
this section depend on the particular choice of the operator.

\begin{figure}
\centerline{
\epsfysize 2.0in
\epsfbox{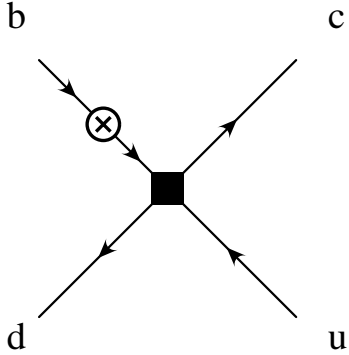}}
\vskip0.5cm
\caption{Same as Fig.~\ref{fig:fig1} but with the electromagnetic
current coupling to the $b$-quark.}
\label{fig:fig2}
\end{figure}

We will now argue that for heavy $b$ and $c$ quarks the non-local
matrix element in Eq.~(\ref{T-prod}) is well approximated by the matrix
element of a sum of local operators. The approximation is valid
provided $\LQCD\ll m_{c,b}$, \ie, the corrections are order $\LQCD/
m_{c,b}$. There are also corrections of order $\LQCD m_{b,c}/q^2$. So
our results are limited to the region were $q^2$ scales like
$m_{c,b}^2$. The region were $q^2$ does not scale like $m_{c,b}^2$ is
parametrically small, so the arguments we present are theoretically
sound. However, there is the practical issue of determining a minimum
$q^2$ for realistic calculations were our approximations can still be
trusted. We return to this practical matter below, when we attempt to
estimate the rate for this decay.

\begin{figure}
\centerline{
\epsfysize 2.0in
\epsfbox{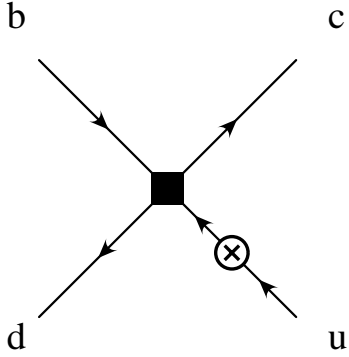}}
\vskip0.5cm
\caption{Same as Fig.~\ref{fig:fig1} but with the electromagnetic
current coupling to the $u$-quark.}
\label{fig:fig3}
\end{figure}

\begin{figure}
\centerline{
\epsfysize 2.0in
\epsfbox{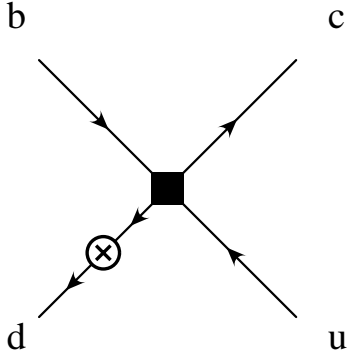}}
\vskip0.5cm
\caption{Same as Fig.~\ref{fig:fig1} but with the electromagnetic
current coupling to the $d$-quark.}
\label{fig:fig4}
\end{figure}

The underlying decay is represented in the quark diagrams of
Figs.~\ref{fig:fig1}--\ref{fig:fig4}. In the heavy quark limit,
$\LQCD\ll m_{c,b}$, the heavy meson momentum is predominantly the
heavy quark's. This suggests the following kinematics in the quark
diagrams: for the momenta of the heavy quarks take $m_bv+k_b$ and
$m_cv'+k_c$, for the momenta of the light quarks take $k_u$ and $k_d$
and then the photon's momentum is determined by conservation,
$q=m_bv-m_cv'+\sum k_i$. We can now exhibit our OPE by considering
the quark Green functions in Figs.~\ref{fig:fig1}--\ref{fig:fig4}. The
convergence of the expansion for physical matrix elements rests on the
intuitive fact that the residual momenta $k_i$ will be of order
$\LQCD$ (parametrically all we need is that these are independent of the
large masses). This intuition is made explicit in  Heavy Quark Effective
Theory (HQET): there are no heavy masses in the HQET-Lagrangian so the
only relevant dynamical scale is $\LQCD$. Thus our expansion of a
non-local product will be in terms of local operators of the HQET.

Calculating the Feynman diagram of Fig.~\ref{fig:fig1} with our choice of
kinematics  we have
\beq
\label{eq:feynmfull1}
-iQ_c\gamma^\mu\frac{i}{\sla{q}+m_c \sla{v}'+\sla{k}_c-m_c}\gamma^\nu P_-
\otimes \gamma_\nu P_-.
\eeq
Here $Q_c=2/3$ is the charge of the $c$-quark and the tensor product
corresponds to the two fermion bilinears. External legs are
amputated. Using $q=m_bv-m_cv'+\sum k_i$ and expanding in
$k_i/m_{c,b}$ we obtain, to leading order
\beq
\label{eq:feynmeff1}
Q_c\gamma^\mu\frac{m_b \sla{v}+m_c}{m_b^2-m_c^2}\gamma^\nu P_-
\otimes \gamma_\nu P_-.
\eeq
This Green's function is that of a local operator in the
HQET. Denoting by $h_v^{(Q)}$ the annihilation operator for the heavy
quark with four-velocity $v$, we define
\beq
\label{eq:hqetopdefd}
\tilde{\cal O}\equiv \bar d \Gamma^{\phantom{()}}_bh^{( b)}_{v}\;
 \bar h^{(c)}_{v'}\Gamma^{\phantom{()}}_c u.
\eeq
Here $\Gamma_{b,c}$ are arbitrary Dirac matrices. With
$\Gamma_c\otimes \Gamma_b$ set equal to the tensor product in
(\ref{eq:feynmeff1}),
\beq
\label{eq:feynmeff1b}
\Gamma_c\otimes \Gamma_b =
Q_c\gamma^\mu\frac{m_b \sla{v}+m_c}{m_b^2-m_c^2}\gamma^\nu P_-
\otimes \gamma_\nu P_-,
\eeq
the operator expansion is
\beq
\label{eq:OPE1}
\int d^4x\;e^{iq\cdot x}
\; T[\bar c\gamma^\mu c(x)\;{\cal O}(0)] = \tilde{\cal O}+\cdots.
\eeq
The ellipses indicate terms of higher order in our expansion, and
correspond to higher derivative operators suppressed by powers of
$m_{c,b}$. There are also perturbative corrections to this
expression. These show up as modifications to the operator defined by 
setting $\Gamma_c\otimes \Gamma_b$ equal to (\ref{eq:feynmeff1}). 

The diagram of Fig.~\ref{fig:fig2} can be analyzed in complete
analogy. It leads to the operator $\tilde{\cal O}$ with the choice
\beq
\label{eq:feynmeff2}
\Gamma_c\otimes \Gamma_b =-
Q_b \gamma^\nu P_- \otimes 
\gamma_\nu P_-\frac{m_b +m_c\sla{v}'}{m_b^2-m_c^2} \gamma^\mu,
\eeq
where $Q_b=-1/3$ is the $b$ quark charge.

The analysis of Figs.~\ref{fig:fig3} and~\ref{fig:fig4} is similar,
but there is an important distinction. With the electromagnetic current
coupling to the light quarks, we get  intermediate light quark
propagators. The denominator in these propagators are parametrically
large only if $q^2$ is parametrically large, \ie, if $q^2\sim
m_{c,b}^2$. With this caveat, the OPE for Fig.~\ref{fig:fig3} gives
\beq
\label{eq:feynmeff3}
\Gamma_c\otimes \Gamma_b =
-Q_u 
\gamma^\nu P_-\frac{\sla{q}}{q^2} \gamma^\mu
\otimes \gamma_\nu P_- 
\eeq
and for Fig.~\ref{fig:fig4} the OPE gives
\beq
\label{eq:feynmeff4}
\Gamma_c\otimes \Gamma_b =
Q_d 
\gamma^\nu P_-\otimes \gamma^\mu \frac{\sla{q}}{q^2} 
\gamma_\nu P_-.
\eeq

\section{Spin Symmetry}
\label{sec:spinsym}
We have shown how to replace the time ordered product in
Eq.~(\ref{T-prod}) by a local operator. The replacement is valid
provided the invariant mass of the lepton pair is large, \ie, scales
as $q^2\sim m_{c,b}^2$. The operator $\tilde {\cal O}$ that replaces
the time ordered product is defined by Eq.~(\ref{eq:hqetopdefd}), with
the tensor $\Gamma_c\otimes \Gamma_b $ defined as the sum of the
contributions in Eqs.~(\ref{eq:feynmeff1b}), (\ref{eq:feynmeff2}),
(\ref{eq:feynmeff3}) and~(\ref{eq:feynmeff4}). We now show how to
relate the matrix element of this operator to the operator with 
$\Gamma_c\otimes \Gamma_b =\gamma^\nu P_-\otimes \gamma_\nu P_-$. This
operator is not only simpler, but one can estimate its matrix elements
by a variety of means, as we explain below. 

Consider the matrix element of $\tilde{\cal O}$ as defined in
Eq.~(\ref{eq:hqetopdefd}) for arbitrary tensor product
$\Gamma_c\otimes \Gamma_b$ between heavy meson states. We will use
heavy quark spin symmetry to determine the matrix elements of this
operator between heavy meson states. Recall that the HQET lagrangian
\beq 
{\cal L}_{\rm HQET}=\barhbv iv\cdot D\hbv
+ \bar h^{(c)}_{v'} iv'\cdot Dh^{(c)}_{v'} 
\eeq 
is symmetric under the group $SU(2)_b\times SU(2)_c$ of
transformations acting on spin indices of the heavy quark fields:
\[
\hbv \to S^{\phantom{\dagger}}_b\hbv
\quad,\quad h^{(c)}_{v'}  \to S^{\phantom{\dagger}}_ch^{(c)}_{v'}.
\]
At $v'=v$ the symmetry is enlarged to $U(4)$, which contains an
$SU(2)$ subgroup corresponding to a flavor symmetry. For now we will
need only the spin symmetries. 

In order to make use of these symmetries, it is convenient to
represent a spin multiplet consisting of a pseudoscalar $P$ and a
vector meson $V_\mu$ by a $4\times4$ matrix
\beq
H_v=\left(\frac{1+\sla{v}}{2}\right)[V_\mu\gamma^\mu-P\gamma_5].
\eeq
Then  $S_b\in SU(2)_b$ and $S_c\in SU(2)_c$ act simply on the left,
\beq
H_v^{(b)}\to S_b H_v^{(b)} \qquad H_{v'}^{(c)}\to S_c H_{v'}^{(c)},
\eeq
while an arbitrary rotation $R$ represented by the Dirac matrix ${\cal
D}(R)$ acts simultaneously on both multiplets according to
\beq
H^{(Q)}\to {\cal D}(R)^\dagger H^{(Q)} {\cal D}(R).
\eeq
Consider now the matrix element $\vev{H^{(c)}_{v'}|\tilde{\cal
O}|H^{(b)}_v}$. It must be linear in the tensors $\Gamma_c\otimes
\Gamma_b$, $H^{(b)}_v$ and $H^{(c)}_{v'}$. Acting with $SU(2)_b$ we
see that $\Gamma_b\to \Gamma_b S_b^\dagger$ and $H_v^{(b)}\to S_b
H_v^{(b)}$, so they enter the matrix element as the product
$\Gamma_bH_v^{(b)}$. A similar argument with $SU(2)_c$ gives then
\beq
\label{eq:fromspin}
 \vev{H^{(c)}_{v'}|\tilde{\cal O}|H^{(b)}_v} \propto
\bar H^{(c)}_{v'} \Gamma_c\otimes \Gamma_b H^{(b)}_v
\eeq
Finally, invariance under rotations implies that the remaining four
indices must be contracted. There are two possible contractions,
\beq
\label{eq:contractions}
{\rm Tr}(\bar H^{(c)}_{v'} \Gamma_c) {\rm Tr} (\Gamma_b H^{(b)}_v)
\quad\hbox{and}\quad 
{\rm Tr}(\bar H^{(c)}_{v'} \Gamma_c \Gamma_b H^{(b)}_v).
\eeq
We now show that the second one is excluded by chiral symmetry. The lagrangian
for a  massless quark in QCD,
\beq
{\cal L}=\bar\psi i\Sla{D}\psi,
\eeq
is invariant under the chiral symmetry
\beq
\psi \to e^{i\alpha\gamma_5}\psi,
\eeq
where $\alpha$, the  parameter of the transformation, is a real number.
Under this symmetry the transformation rule for our tensors is
\beq
\label{eq:GbHbtransf}
\Gamma_b H^{(b)}_v \to e^{-i\alpha\gamma_5} 
\Gamma_b H^{(b)}_v e^{i\alpha\gamma_5}
\eeq
and
\beq
\bar H^{(c)}_{v'} \Gamma_c \to 
e^{i\alpha\gamma_5}\bar H^{(c)}_{v'} \Gamma_c  e^{-i\alpha\gamma_5}.
\eeq
It is seen that the first contraction of indices in
(\ref{eq:contractions}) is invariant, but the second one is not.

We have shown that heavy quark spin symmetry, rotations and light quark chiral
symmetry combine to give
\beq
\label{eq:werels}
\vev{H^{(c)}_{v'}|\tilde{\cal O}|H^{(b)}_v} =\frac14\beta(w)
{\rm Tr}(\bar H^{(c)}_{v'} \Gamma_c) {\rm Tr} (\Gamma_b H^{(b)}_v).
\eeq

We have indicated that the invariant matrix element $\beta$ is a
function of $w=v\cdot v'$. In general, it is a function  of $v$
and $v'$. However, since it must be Lorenz invariant and since
$v^2=v^{\prime2}=1$, it is a function of $w=v\cdot v'$ only.

The octet operator in the HQET, 
\beq
\label{eq:hqetopdefd8}
\tilde{\cal O}_8\equiv \bar d \Gamma^{\phantom{()}}_bT^ah^{(b)}_{v}\;
 \bar h^{(c)}_{v'}\Gamma^{\phantom{()}}_cT^a u, 
\eeq has the same spin
and heavy flavor symmetry properties as its singlet
counterpart. Therefore in complete analogy we can introduce a reduced
matrix element $\beta_8$:
\beq
\label{eq:werels8}
\vev{H^{(c)}_{v'}|\tilde{\cal O}_8|H^{(b)}_v} =\frac14\beta_8(w)
{\rm Tr}(\bar H^{(c)}_{v'} \Gamma_c) {\rm Tr} (\Gamma_b H^{(b)}_v).
\eeq

The authors of Ref.~\cite{GJMSW} proposed a relation analogous to
Eq.~(\ref{eq:werels}) for a $\Delta B=2$ transition. It was noted
there that spin symmetry allowed more than one invariant and that,
however, all invariants lead to the same symmetry relations. One may
wonder if our use  of chiral symmetry may help relate the different
invariants there. We show that this is not the case. For the $\Delta
B=2$ case the analogue of Eq.~(\ref{eq:fromspin}) is
\beq
\label{eq:fromspin2}
 \vev{H^{(\bar b)}_{v}|\tilde{\cal O}_{\Delta B=2}|H^{(b)}_v} \propto
 \Gamma_{\bar b}\bar H^{(\bar b)}_{v}\otimes \Gamma_b H^{(b)}_v,
\eeq
where $\tilde{\cal O}_{\Delta B=2} =
\bar d \Gamma^{\phantom{()}}_{\bar b}h^{(\bar b)}_{v}\;
\bar d \Gamma^{\phantom{()}}_bh^{( b)}_{v}$ (note that we define
$h^{(\bar b)}_{v}$ to create a $b$-antiquark). Again,
invariance under rotations implies that the remaining four
indices must be contracted and, again, there are two possible contractions,
\beq
\label{eq:contractions2}
{\rm Tr}(\Gamma_{\bar b}\bar H^{(\bar b)}_{v}) {\rm Tr} (\Gamma_b H^{(b)}_v)
\quad\hbox{and}\quad 
{\rm Tr}(\Gamma_{\bar b}\bar H^{(\bar b)}_{v}\Gamma_b H^{(b)}_v).
\eeq
Chiral symmetry for the antiquark's meson tensor is just as for the
quark's in Eq.~(\ref{eq:GbHbtransf}),
\beq
\Gamma_{\bar b}\bar H^{(\bar b)}_{v}\to 
e^{-i\alpha\gamma_5} \Gamma_{\bar b}\bar H^{(\bar b)}_{v} e^{i\alpha\gamma_5}.
\eeq
Therefore both contractions in (\ref{eq:contractions2}) are allowed by
chiral symmetry. However, it is easy to see that for a class of operators of
interest the two contractions are equivalent. If
\[
\Gamma_{\bar b}\otimes \Gamma_b = \gamma^\mu P_- \hat\Gamma \otimes
\gamma_\mu P_-
\] 
or
\[
\Gamma_{\bar b}\otimes \Gamma_b = \gamma^\mu P_-  \otimes
\gamma_\mu P_-\hat\Gamma ,
\] 
for any arbitrary Dirac matrix $\hat\Gamma$ the two contractions are
related by Fierz rearrangement. This class of operators
includes the $B-\bar B$ mixing case studied in Ref.~\cite{GJMSW}.

\section{QCD Corrections}
\label{sec:QCD1}
Consider the operator expansion in Eq.~(\ref{eq:OPE1}). We have seen
that at leading order the operator on the right hand side is given by
Eqs.~(\ref{eq:hqetopdefd})--(\ref{eq:feynmeff1b}). We now consider the
leading-log corrections to this relation. In the large mass limit
these are formally the largest, leading corrections to the operator
expansion. 
A renormalization scale
$\mu$ must be stipulated for the evaluation of matrix elements of the
composite operators on both sides of Eq.~(\ref{eq:OPE1}). It is often
convenient to evaluate the matrix elements at a low renormalization
point $\mu=\mu_{\rm low}$. This choice makes the matrix elements
in the HQET completely independent of the large masses of the heavy
quarks. If $\mu_{\rm low}\ll m_{c,b}$ there are large corrections to 
Eq.~(\ref{eq:OPE1}) in the form of powers of
$\alpha_s\ln(m_{c,b}/\mu_{\rm low})$. These powers of large logarithms
can be summed  using renormalization group techniques. The corrections
to these ``leading-logs'' are of order $1/\ln(m_{c,b}/\mu_{\rm low})$
or $\alpha_s$. It is important therefore to keep $\mu_{\rm low}$
small, but large enough that perturbation theory remains valid. When
we estimate decay rates below, we use $\mu_{\rm low}=1.0$~GeV.

To study the dependence on the renormalization point $\mu$ we take a
logarithmic derivative on both sides of Eq.~(\ref{eq:OPE1}). Consider
first the left side. Acting
with $\mu(d/d\mu)$ on the charm
number current $\bar c \gamma^\mu c$ gives zero, because the current
is conserved. The action of  $\mu(d/d\mu)$  on the composite
four-quark operator is a linear combination of  itself and the octet
operator. It is therefore convenient to consider instead the linear
combination that appears in the effective Hamiltonian (\ref{eq:Heff}):
\begin{eqnarray}
\label{eq:OPEfull}
\int d^4x\;e^{iq\cdot x}
\; T[\bar c\gamma^\mu c(x)& &(c{\cal O}(0)+c_8{\cal O}_8(0))] =\nonumber\\ 
& &\tilde c\tilde{\cal O}+\tilde c_8\tilde{\cal O}_8+\cdots.
\end{eqnarray}
The coefficients $c$ and $c_8$ are such that the left hand side is
$\mu$-independent. This is necessary for the physical amplitude to be
independent of the arbitrary choice of renormalization point $\mu$. 
Therefore our task is to determine the proper $\mu$-dependence for
$\tilde c$ and $\tilde c_8 $  so that the right hand side is also
independent of $\mu$. Therefore, if the operators satisfy
\beq
\label{eq:rgeops}
\mu\frac{d}{d\mu}
\left(
\begin{array}{c}
\tilde{\cal O} \\
\tilde{\cal O}_8 \\
\end{array}\right)
=\gamma \left(
\begin{array}{c}
\tilde{\cal O} \\
\tilde{\cal O}_8 \\
\end{array}\right),
\eeq
where $\gamma$ is a $2\times2$ matrix of anomalous dimensions, then
the coefficients must satisfy
\beq
\label{eq:rgecoefs}
\mu\frac{d}{d\mu}
\left(
\begin{array}{c}
\tilde{c} \\
\tilde{c}_8 \\
\end{array}\right)
=-\gamma^T \left(
\begin{array}{c}
\tilde{c} \\
\tilde{c}_8 \\
\end{array}\right).
\eeq
Here ``$T$'' denotes transpose of a matrix.

\begin{figure}
\centerline{
\epsfysize 4.0in
\epsfbox{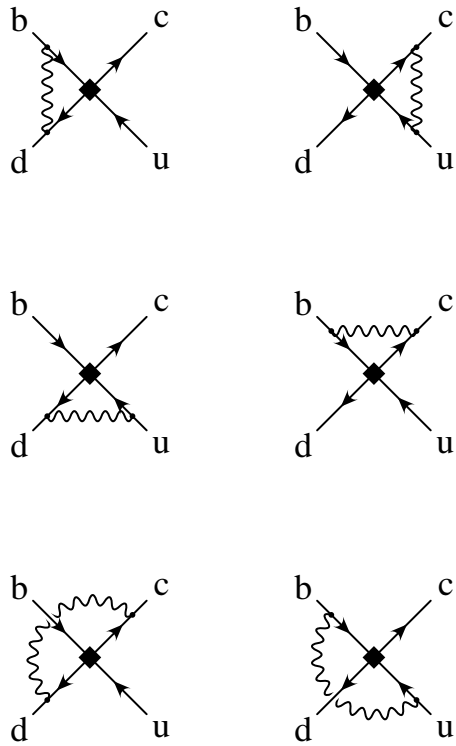}}
\vskip0.5cm
\caption{One loop Feynman diagrams for the calculation of the
anomalous dimension matrix. The solid diamond represents the local
operators ${\cal O}$ or ${\cal O}_8$.}
\label{fig:loops}
\end{figure}

The calculation of the anomalous dimension matrix is
straightforward. In dimensional regularization with $D=4-\epsilon$
dimensions, one needs\cite{DG} the residues of the $\epsilon$-poles of
graphs with one insertion of the operators $\tilde{\cal O}$ and
$\tilde{\cal O}_8 $. The leading-log corrections arise from the
leading, $O(\alpha_s)$ terms in $\gamma$. These arise from the one-loop
graphs in Fig.~\ref{fig:loops}

In principle the different tensor structures $\Gamma_c\otimes
\Gamma_b$ defining $\tilde{\cal O}$ and $\tilde{\cal O}_8 $ can have
different anomalous dimensions and even mix among themselves.  However
spin symmetry ensures that the anomalous dimension matrix is
independent of the tensor structure $\Gamma_c\otimes \Gamma_b$.

We find
\begin{equation}
\gamma=\frac{\alpha_s}{4\pi}
\left(\begin{array}{cc}
8 & -4wr(w)-2\\
-\frac89 wr(w)-\frac49 & \frac{17}3-\frac{14}3wr(w))\\
\end{array}\right),
\end{equation}
where
\beq
r(w)\equiv \frac1{\sqrt{w^2-1}}\ln(w+\sqrt{w^2-1}).
\eeq
The solution to the renormalization group equation (\ref{eq:rgecoefs})
is straightforward. In terms of the ratio of running coupling constants
\beq 
\label{eq:zdef}
z\equiv\left(\frac{\alpha_s(\mu)}{\alpha_s({\muz})}\right)
\eeq
and the functions
\begin{eqnarray}
\psi &=& \frac1{12}\;\frac{41-14wr(w)}{b_0},\\
\xi  &=& -\frac34\; \frac{1+2wr(w)}{b_0},
\end{eqnarray}
where the coefficient of the one loop term of the $\beta$-function for
QCD is $b_0=11-\frac23n_f$, and $n_f$ is the number of light flavors
($n_f=3$ in our case), we obtain
\beq
\label{eq:rgesol1}
\left(
\begin{array}{c}
\tilde{c}(\mu) \\
\tilde{c}_8(\mu) \\
\end{array}\right)
=U
\left(
\begin{array}{c}
\tilde{c}({\muz}) \\
\tilde{c}_8({\muz}) \\
\end{array}\right)
\eeq
where
\beq
\label{eq:rgesol2}
U=z^\psi
\left(\begin{array}{cc}
\frac19z^\xi+\frac89z^{-\xi}& \frac4{27}(z^\xi-z^{-\xi})\\
\frac23(z^\xi-z^{-\xi})& \frac89z^\xi+\frac19z^{-\xi}
\end{array}\right).
\eeq

The question that remains is how to determine the coefficients $\tilde
c$ and $\tilde c_8$ at some scale ${\muz}$. But we have already
determined these coefficients in Sec.~\ref{sec:method}. Recall that
the operator $\tilde {\cal O}$ that replaces the time ordered product
is defined by Eq.~(\ref{eq:hqetopdefd}), with the tensor
$\Gamma_c\otimes \Gamma_b $ defined as the sum of the contributions in
Eqs.~(\ref{eq:feynmeff1b}), (\ref{eq:feynmeff2}),
(\ref{eq:feynmeff3}) and~(\ref{eq:feynmeff4}) with unit coefficient.
The question can be rephrased as what is the scale ${\muz}$ for which
the calculation in Sec.~\ref{sec:method} is valid. What we would like
to do is to determine for what choice of ${\muz}$  the loop
corrections to relations like Eq.~(\ref{eq:OPE1}) will be free from large
logs.  The only relevant scales in the problem are the large masses
$m_{c,b}$, the invariant mass of the $e^+e^-$ pair, $q^2$, which
itself scales like $m_{c,b}^2$, the small masses and residual momenta
and the renormalization point ${\muz}$. The corrections to the
relations of Sec.~\ref{sec:method} are guaranteed to be free from logs
of the small masses or residual momenta. But there will be logs of
ratios of large masses to the renormalization point,
$\ln(m_{c,b}/{\muz})$. To avoid these one may choose ${\muz}\sim
m_{c,b}$. For our computations below we will use
${\muz}\approx4.0$~GeV. If the scales $m_c$ and $m_b$ are both large
but very disparate one could review the above analysis by introducing
a new renormalization group equation to re-sum the logs of
$m_c/m_b$. The results of this section would still  re-sum the logs of
$\mu/m_c$.

We thus have that $\tilde c(\mu)$ and $\tilde c_8(\mu)$ are given by
Eqs.~(\ref{eq:rgesol1}) and~(\ref{eq:rgesol2}), with
\begin{eqnarray}
\tilde c(\muz) & = & c({\muz}) = \frac23(x^{-1}-\frac12x^2)\\
\tilde c_8({\muz}) & = & c_8({\muz}) = x^{-1}+x^2
\end{eqnarray}
where
\beq
x\equiv\left(\frac{\alpha({\muz})}{\alpha(M_W)}\right)^{6/23}.
\eeq

For illustration we have given the leading log expression for the
coefficients $c(\muz)$ and $c_8({\muz})$, but in rate computations
below we use the next to leading log results from \cite{buras}.  We do
not have at present a full next to leading log result: still missing is a
computation of the one loop corrections to the coefficients $\tilde c$
and $\tilde c_8$ at $\mu=\muz$ and of the anomalous dimensions matrix
$\gamma$ of Eq.~(\ref{eq:rgeops}) 
at two loops. It is interesting to note that the coefficients
$c(\mu_0)$ are significantly enhanced at next to leading log
order. For the case $\mu_0=4.0$~GeV one has in next to leading
order\cite{buras} $c=0.16$, rather than the leading log result
$c=0.07$. We emphasize that this
enhancement can be systematically accounted for. The large enhancement
is not a signal of perturbation theory breaking down but rather due to
the accidental cancellation in the leading order.

\section{Rates: $\bar B^0\to D^{(*)0} e^+e^-$}
\label{sec:results1}
We are ready to compute  decay rates. Defining
\beq
h^{(*)\mu}=
\langle D^{(*)0}| \int d^4x\;e^{iq\cdot x}
\; T(j^\mu_{\rm em}(x){\cal H}'_{\rm eff}(0)) |B^0\rangle,
\eeq
the decay rate for $B^0\to D^{(*)0}e^+e^-$ is given in terms of $q^2$
and $t\equiv(p_D+p_{e^+})^2=(p_B-p_{e^-})^2$ by   
\begin{equation}
\label{eq:doublediffrate}
\frac{d\Gamma}{dq^2dt}=\frac1{2^8\pi^3M_B^3}
\left|\frac{e^2}{q^2}\ell_\mu h^{(*)\mu} \right|^2
\end{equation}
where $\ell^\mu=\bar u(p_{e^-})\gamma^\mu v(p_{e^+})$ is the leptons'
electromagnetic current. A sum over final state lepton helicities, and
polarizations in the $D^*$ case, is implicit.

To compute $h^{(*)\mu}$ we need to pull together the results of the
previous sections.  First the time ordered product is expanded in
terms of local operators as in
Eqs.~(\ref{eq:feynmeff1b})--(\ref{eq:feynmeff4}). This involves
replacing the coefficient functions $c({\muz})$ and $c_8({\muz})$ by
$\tilde c({\muz})$ and $\tilde c_8({\muz})$ as seen in
Eq.~(\ref{eq:OPEfull}). Then the matrix elements of the leading local
operators $\tilde{\cal O}$ and $\tilde{\cal O}_8$ between particular
states can all be expressed in terms of the reduced matrix elements
$\beta$ and $\beta_8$ defined in (\ref{eq:werels}) and
(\ref{eq:werels8}). Finally, to make all dependence on the heavy quark
masses explicit, we run down the coefficients $\tilde c$ and $\tilde
c_8$ from the scale $\mu={\muz}$ of order of $m_{b,c}$ (which we
take to be $\sqrt{m_cm_b}$) to a scale $\mu=\mu_{\rm low}$ of order of
a few times $\LQCD$. 

Our computation gives 
\begin{eqnarray}
h^\mu=\frac\kappa3& &\Big[
\frac{-(2wm_b+m_c)v^\mu-(m_b-4wm_c)v^{\prime\mu}}{(m_bv-m_cv')^2}
\nonumber\\
& &\hspace{3cm}+\frac{3(m_bv^{\prime\mu}+m_cv^\mu)}{m_b^2-m_c^2}\Big]
\end{eqnarray}
and 
\begin{eqnarray}
h^{*\mu}&=&\frac\kappa3\Big[
\frac{m_b(\epsilon^\mu+2v\cdot\epsilon v^{\mu})
-m_c(3v\cdot\epsilon v^{\prime\mu}+w\epsilon^\mu)}{(m_bv-m_cv')^2}\nonumber\\
& &\hspace{2cm}+
\frac{3im_c\epsilon^{\mu\alpha\beta\gamma}\epsilon_\alpha v'_\beta v_\gamma}%
{(m_bv-m_cv')^2}
\\
&-& \frac{3m_b\epsilon^\mu+3m_c(v\cdot\epsilon
v^{\prime\mu}-w\epsilon^\mu)
-im_c\epsilon^{\mu\alpha\beta\gamma}\epsilon_\alpha v'_\beta v_\gamma}%
{m_b^2-m_c^2}
\Big].\nonumber
\end{eqnarray}
Here $\kappa=G_F/\sqrt2\,V^{\phantom{*}}_{cb}V_{ud}^*[\tilde c\beta+\tilde
c_8\beta_8]$. These expressions are our central results, demonstrating
that the decay rates for $B^0\to D^{(*)0}e^+e^-$ can be expressed in
terms of the matrix elements $\beta$ and $\beta_8$. Below we make an
educated guess of these matrix elements, but for reliable results they
should be determined from first principles, say, by Monte Carlo
simulations of lattice QCD.

In the computation of the rate the amplitude depends on heavy quark
masses $m_c$ and $m_b$, while the phase space involves physical meson
masses $M_B$ and $M_{D}$ or $M_{D^*}$. Although it is straightforward
to retain the dependence on all four masses in our expressions for the
decay rates, we have chosen to express the results in terms of
physical meson masses, with the substitutions $m_b=M_B$ and
$m_c=M_{D}$ or $m_c=M_{D^*}$. We are not justified in distinguishing
between quark and meson masses since the distinction enters at higher
order in the $1/m_{c,b}$ expansion.

It is now a trivial exercise to compute the differential decay
rate. Integrating the rate in Eq.~(\ref{eq:doublediffrate}) over the
variable $t$ we obtain
\beq
\label{eq:rateBDee}
\frac{d\Gamma}{dq^2}=
\frac{\alpha^2 G_F^2}{288\pi M_B^3}|V_{cb} V_{ud}|^2 
(\tilde c\beta+\tilde c_8\beta_8)^2{\cal F}(\hq).
\eeq
Here ${\cal F}(\hq)$ is a dimensionless function of $\hq \equiv
\sqrt{q^2/m_b^2}$ and $\hm\equiv M_{D^{(*)}}/M_B$. For $B^0\to
D^{*0}e^+e^-$ it is given by
\begin{eqnarray}
{\cal F} &=&
\frac43 \frac{\sqrt {{{1-2 {\hat q}^{2}-2 {\hat m}^{2}+{\hat q}^{4}-2 {\hat m}
^{2}{\hat q}^{2}+{\hat m}^{4}}} }}{{\hat q}^6\hat m (1-\hat m^2)^{2}} 
 \nonumber\\ 
& &(5 {\hm}^{2}+19 {\hm}^{4}\hq^{2}+30 {\hm}^{6}-20 {\hm}^{4}-14 \hq^{2}
{\hm}^{2}\nonumber\\ 
& &-20 {\hm}^{8}+12 \hq^{6}{\hm}^{2}+\hq^{2}+\hq^{6}-2 \hq^{4}+2 {\hm}^{6}\hq^{4}\nonumber\\ 
& &-6 {\hm}^{8}\hq^{2}+5 {\hm}^{10}-6 \hq^{6}{\hm}^{4}
+5 {\hm}^{2}\hq^{8}),\nonumber\\
\end{eqnarray}
while for $B^0\to D^0e^+e^-$
\beq
{\cal F}=
\frac{4   (2 {\hat m}^{2}+1  )^{2}  (1-2 {\hat q}^{2}-2
 {\hat m}^{2}+{\hat q}^{4}-2 {\hat m}^{2}{\hat q}^{2}+{\hat m}^{4})^{\frac32}}{
3{\hat q}^4\hat m (1-\hat m^2)^{2} }.
\eeq
In these we have neglected the electron mass.

In order to obtain a numerical estimate of the branching fraction we
need to calculate the hadronic matrix elements $\beta$ and
$\beta_8$. While these could be studied in Monte Carlo simulations of
QCD on the lattice, at the moment we have no reliable information on
their magnitude. These matrix elements are similar to the matrix
element of the $\Delta B=2$ operator for $B-\bar B$ mixing. Lattice
QCD\cite{latticeBB} indicates that the vacuum saturation approximation
works very well for $B-\bar B$ mixing. Therefore we take vacuum
saturation as an educated guess\footnote{The matrix elements in
$B^+\to D^{(*)+} e^+e^-$ can be related by symmetry to the matrix
element for $B-\bar B$ mixing, if the matrix element of the octet is
negligible; see Ref.~\cite{evans-99-1}} for $\beta$ and~$\beta_8$. Taking
$\Gamma_c\otimes\Gamma_b=\gamma^\mu\gamma_5\otimes\gamma^\nu\gamma_5$
the right hand side of Eq.~(\ref{eq:werels}) is $v^\nu
v^{\prime\mu}\beta$. On the left hand side vacuum saturation gives
$(z^{-a_I}f_Bp_B^\nu/\sqrt{M_B})(z^{-a_I}f_Dp_D^\mu/\sqrt{M_D})$. Here $z$ is
defined in Eq.~(\ref{eq:zdef}) and $a_I=2/b_0$\cite{fggw} is the well
known anomalous scaling power for the heavy-light current in
HQET.\footnote{The two factors of $z^{-a_I}$ really correspond to
distinct running, between $m_b$ and $\mu_{\rm low}$ for the first
factor, and between $m_c$ and $\mu_{\rm low}$ for the second. The
distinction is of higher order than we have retained, if we assume
that the heavy scales $m_b$ and $m_c$ are not too disparate, that is,
that $\alpha_s$ does not run much between these scales.} Thus we
obtain
\begin{eqnarray}
\beta(w) & = & z^{-2a_I} f_Bf_D\sqrt{M_BM_D}\\
\beta_8(w) & = &0
\end{eqnarray}
The second equation is true not just in vacuum saturation but also in
the approximation that we can insert a complete set of states
between the currents defining $\tilde {\cal O}_8$. This is not an
exact statement because the composite operator $\tilde {\cal O}_8$
does not equal the product of two currents. But the distinction arises
from their different short distance behavior. So we expect the
deviation of $\beta_8$ from zero to be of order of the QCD coupling at
short distances $\alpha(\mu_0)$ times the unsuppressed $\beta$. 

Using these matrix elements we integrate the differential rate in
Eq.~(\ref{eq:rateBDee}) over the range $1.0~\hbox{GeV}\le q^2\le
q^2_{\rm max}$ to obtain a partial decay rate. We have chosen
$q^2_{\rm min}=1.0$~GeV as a lower limit since our OPE requires that
$q^2$ scale like $m_{c,b}^2$. The corrections to the leading terms in
Eqs.~(\ref{eq:feynmeff3}) and~(\ref{eq:feynmeff4}) are  of the form of
an expansion in $m_{c,b}k/q^2$, where $k$ is any of the residual
momenta and in our matrix elements is of order
$\LQCD$. Parametrically, if $q^2\sim m_{c,b}^2$, then $m_{c,b}k/q^2
\sim\LQCD/m_{c,b}\ll1$. In addition, the region over which
$q^2\alt\LQCD m_{c,b}$ where the expansion breaks down, is
parametrically small. However, physical heavy masses are not very
large, and the scale $m_b\LQCD$ is just slightly smaller than
$m_c^2$. In order to have some non-trivial phase space we have taken
$q^2\agt m_b\LQCD\sim1.0$~GeV. The price we pay is that for the lower
values of $q^2$ our expansion converges slowly, $m_{c,b}k/q^2\alt1$.

We find
\begin{eqnarray}
\label{eq:ratedstar}
{\rm Br}(B^0\to D^{*0}e^+e^-)|_{q^2>1~{\rm GeV}}
&=&1.4\times10^{-8}\\
{\rm Br}(B^0\to D^{0}e^+e^-)|_{q^2>1~{\rm GeV}}
&=&2.6\times10^{-9}
\end{eqnarray}
where we have used $|V_{cb}V_{ud}|=0.04$, $f_D=f_B\sqrt{M_B/M_D}$ and
$f_B=170$~MeV. It is important to observe that the portion of phase
space $q^2\ge 1.0~{\rm GeV}$ is expected to give a small fraction of
the total rate since the pole at $q^2=0$ dramatically amplifies
the rate for small $q^2$. The rates for $B_s^0\to D^{*0}e^+e^-$ and
$B_s^0\to D^{0}e^+e^-$ can be obtained to good approximation by
replacing $|V_{cb}V_{ud}|$ by $|V_{cb}V_{us}|$, reducing the rates by
$(0.22)^2\approx0.05$.

The next generation of B-physics experiments at 
high energy and luminosity hadron colliders, like LHC-B and BTeV,
will produce well in excess of $10^{11}$ $B$-mesons per year. Our
calculation includes only large invariant mass
lepton pairs so detection and triggering on the lepton pair should be
straightforward. Dedicated studies must be done to determine
feasibility of detection and measurement of spectra of these decays.

\section{Decays to Quarkonium}
\label{sec:onium}
\subsection{Operator Expansion and NRQCD}
The decays $B_s\to\eta_c e^+e^-$ and $B_s\to J/\psi e^+e^-$ (and
obvious extensions to excited charmonium) can be studied in a similar
way. The notable difference in the operator expansion here is that the
residual momenta $k$ of the heavy quarks in the quarkonium bound state
do scale with the large heavy mass $k\sim\alpha_s m_c$, as opposed to
the residual momenta of the quarks in the heavy $B$ or $D$ mesons,
$k\sim\LQCD$. The residual momentum for the case of quarkonia is small
for a different reason: ${\bf k}=m_c{\bf u} $ and $k^0=\frac12m_c u^2$
are small because the velocity ${\bf u}$ of the bound quarks is
small\cite{NRQCD} for heavy quarks, $u\sim \alpha_s(m_c)$. The
parameter of the expansion is therefore $m_{c,b}k/m_{c,b}^2\sim
\alpha_s(m_c)$.

\begin{figure}
\centerline{
\epsfysize 2.0in
\epsfbox{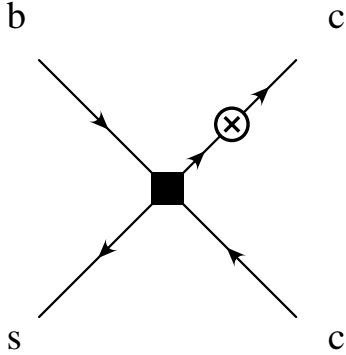}}
\vskip0.5cm
\caption{Feynman diagram representing a contribution to the Green
function. The filled square represents the four quark operator ${\cal
O}$ and the cross represents the electromagnetic current $j^\mu_{{\rm
em}}$, cf. Eq.~(\ref{eq:OPEfullcharmonium}), 
which here couples to the $c$-quark. }
\label{fig:fig5}
\end{figure}

Our best hope in making the nature of the expansion explicit is to use
NRQCD\cite{NRQCD}, the effective theory of non-relativistic quarks in
QCD. As opposed to HQET, where all the heavy mass dependence has
disappeared, the lagrangian of NRQCD still depends on the heavy mass:
\beq
\label{eq:NRQCDlag}
{\cal L}_{\rm NRQCD}= \Psi^\dagger(iD_t -\frac{{\bf D}^2}{2m_c})\Psi
\eeq
Here $\Psi$ denotes a two component spinor field for the $c$-quark. A
separate spinor field must be included to describe the antiquark. We
have written the lagrangian in the rest-frame of charmonium, but it is
straightforward to boost into a moving frame. One relies on the
dynamics to generate the small parameter of the
expansion.\footnote{Attempts to make the expansion in $u$\cite{mlam}
or, alternatively, in $1/c$\cite{bgir} explicit yield theories where
the gluon self-couplings must be perturbative. The scale of QCD must
then be negligible compared with the Bohr radius of quarkonium,
$\LQCD\ll m_c\alpha_s(m_c)$. In our case non-perturbative gluons play a
crucial role in binding the heavy-light meson $B$.} For
example, the two terms in ${\cal L}_{\rm NRQCD}$ are of comparable
magnitude if, as expected, $D_t\sim k^0\sim m_c\alpha_s^2$ and $ |{\bf
D}|\sim |{\bf k}|\sim m_c\alpha_s$.

The operator expansion is in terms of operators with an HQET quark, a
light quark and a pair of NRQCD quark-antiquark. So instead of
Eq.~(\ref{eq:hqetopdefd}) we have
\beq
\label{eq:hqetnrqcdopdefd}
\tilde{\cal O}\equiv \bar d \Gamma^{\phantom{()}}_bh^{( b)}_{v}\;
 \Psi_c^\dagger\Gamma^{\phantom{()}}_c \Psi_{\bar c},
\eeq
where $\Psi_c^\dagger$ and $\Psi_{\bar c}$ create a charm quark and a
charm antiquark, respectively. We elect to use four component spinors
throughout; the reduction to two components results from algebraic
constraints that must be imposed, just as in HQET:
\[
\Psi=\left(\frac{1+\sla{v}'}{2}\right)\Psi
\]

\begin{figure}
\centerline{
\epsfysize 2.0in
\epsfbox{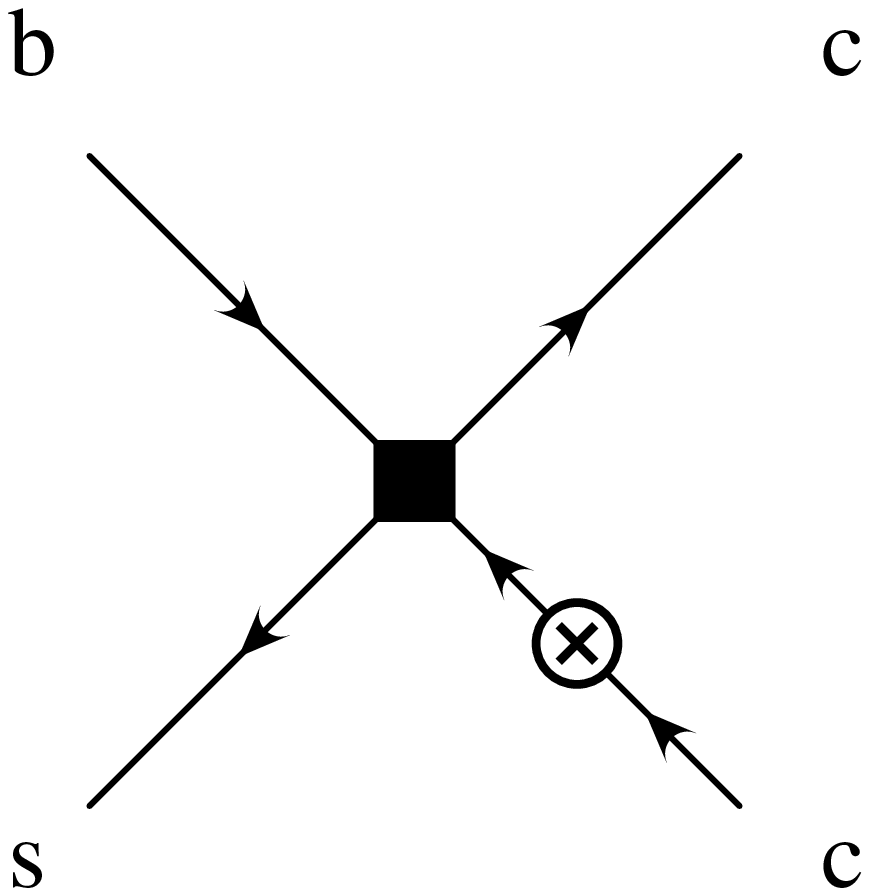}}
\vskip0.5cm
\caption{Same as Fig.~\ref{fig:fig5} but with the electromagnetic
current coupling to the $c$-antiquark.}
\label{fig:fig7}
\end{figure}

\begin{figure}
\centerline{
\epsfysize 2.0in
\epsfbox{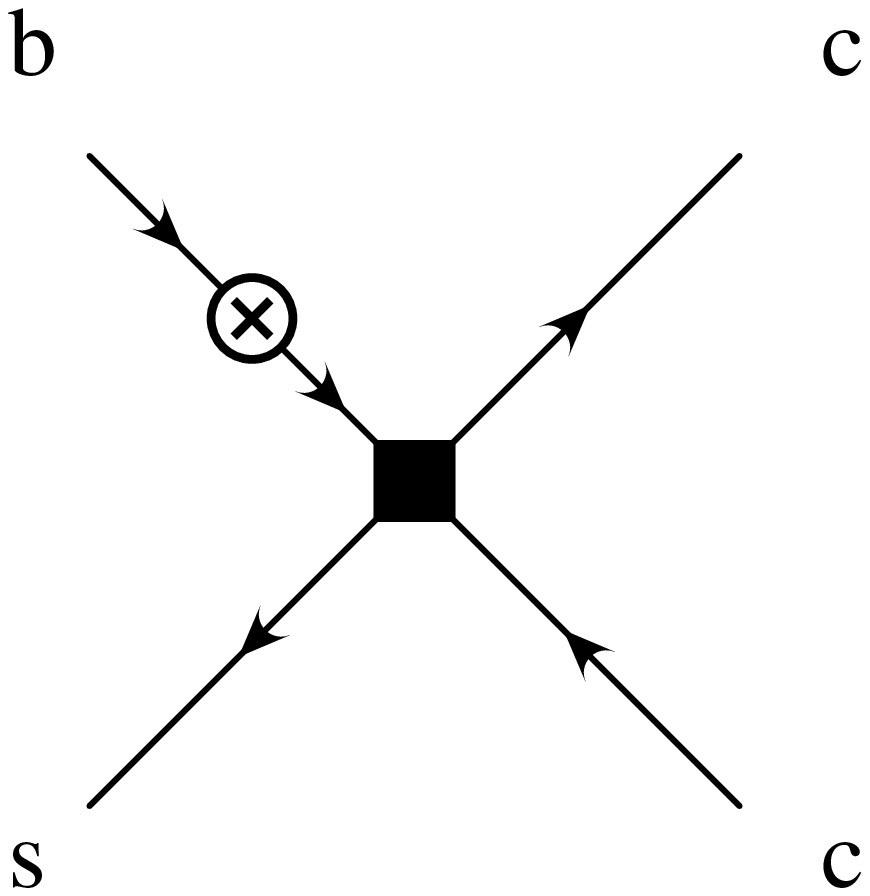}}
\vskip0.5cm
\caption{Same as Fig.~\ref{fig:fig5} but with the electromagnetic
current coupling to the $b$-quark.}
\label{fig:fig6}
\end{figure}

The calculation proceeds much as before. The effective Hamiltonian for
the weak transition is
\beq
\label{eq:Heffcharmonium}
{\cal H}'_{\rm eff}= \frac{4G_F}{\sqrt2}\,V^{\phantom{*}}_{cs}V^*_{cb}\left(
c(\mu/M_W){\cal O}+c_8(\mu/M_W){\cal O}_8\right),
\eeq
where 
\beq
\label{eq:Odefdcharmonium}
{\cal O}=\bar s\gamma^\nu P_- b \;\;\bar c\gamma_\nu P_- c
\eeq
and
\beq
{\cal O}_8=\bar s\gamma^\nu P_-T^a b\;\; 
\bar c\gamma_\nu P_- T^a c.
\eeq
The operator expansion of the
hadronic matrix element takes the form
\begin{eqnarray}
\label{eq:OPEfullcharmonium}
\int d^4x\;e^{iq\cdot x}
\; T[j_{\rm em}^\mu(x) & &(c{\cal O}(0)+c_8{\cal O}_8(0))] =\nonumber\\ 
& &\tilde c\tilde{\cal O}+\tilde c_8\tilde{\cal O}_8+\cdots,
\end{eqnarray}
where $\tilde{\cal O}$ is defined in (\ref{eq:hqetnrqcdopdefd}) and
the octet operator $\tilde{\cal O}_8$ is defined analogously,
\beq
\label{eq:hqetnrqcdopdefd8}
\tilde{\cal O}_8\equiv \bar d \Gamma^{\phantom{()}}_bT^ah^{( b)}_{v}\;
 \Psi_c^\dagger\Gamma^{\phantom{()}}_cT^a \Psi_{\bar c}. 
\eeq
The first task is to determine the tensor
$\Gamma_b\otimes\Gamma_c$. To this order we consider Green functions
of the time ordered product in Eq.~(\ref{eq:OPEfullcharmonium}) with
four external quarks. The in-going momenta of the $b$- and $s$-quarks
are $m_b v+k_b$ and $k_s$, respectively. The outgoing momenta of the
charm pair are $m_c v' +k_c$ and $m_c v' +k_{\bar c}$. As explained
above, we expect $k_b\sim k_s\sim \LQCD$ while $k_c\sim k_{\bar c}\sim
m_c\alpha_s(m_c)$. The leading term in the momentum of the
electromagnetic current is $q=m_bv-2m_cv'$. For the purpose of
determining the expansion coefficients at tree level we may set $c=1$
and $c_8=0$ and, choosing a renormalization point $\muz$ of the order
of the large masses $m_{c,b}$, we can set $\tilde c=1$ and $\tilde
c_8=0$. There are four graphs contributing to the tensor
$\Gamma_c\otimes\Gamma_b$. Fig.~\ref{fig:fig5} gives
\beq
\label{eq:feynmeffch1}
\Gamma_c\otimes\Gamma_b=
Q_c\gamma^\mu\frac{m_b \sla{v}-m_c(\sla{v}'-1)}{m_b^2-2m_bm_cw}\gamma^\nu P_-
\otimes \gamma_\nu P_-,
\eeq
and Fig.~\ref{fig:fig7} gives
\beq
\label{eq:feynmeffch2}
\Gamma_c\otimes\Gamma_b=
Q_c\gamma^\nu P_-\frac{-m_b \sla{v}+m_c(\sla{v}'+1)}{m_b^2-2m_bm_cw}\gamma^\mu
\otimes \gamma_\nu P_-.
\eeq
Note that the denominator, which dictates the convergence of the
expansion, scales with $m_{c,b}^2$. It vanishes at
$w_0=m_b/2m_c$. However, this is never in the physical region: $w_{\rm
max}=(m_b^2+4m_c^2)/4m_bm_c=w_0-(m_b/4m_c-m_c/m_b)$, but $m_b>2m_c$
for the decay to be allowed.

\begin{figure}
\centerline{
\epsfysize 2.0in
\epsfbox{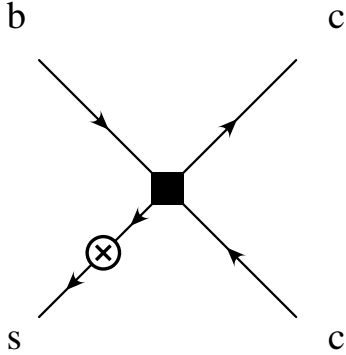}}
\vskip0.5cm
\caption{Same as Fig.~\ref{fig:fig5} but with the electromagnetic
current coupling to the $s$-quark.}
\label{fig:fig8}
\end{figure}

The diagrams in Figs.~\ref{fig:fig6} and~\ref{fig:fig8} are just as in
Figs.~\ref{fig:fig2} and~\ref{fig:fig4}, with the replacement
$q=m_bv-m_cv'\to q=m_bv-2m_cv'$. For the first we have
\beq
\label{eq:feynmeffch3}
\Gamma_c\otimes \Gamma_b =-
Q_b \gamma^\nu P_- \otimes 
\gamma_\nu P_-\frac{m_b +2m_c\sla{v}'}{m_b^2-4m_c^2} \gamma^\mu,
\eeq
and for the second
\beq
\label{eq:feynmeffch4}
\Gamma_c\otimes \Gamma_b =
Q_d 
\gamma^\nu P_-\otimes \gamma^\mu \frac{\sla{q}}{q^2} 
\gamma_\nu P_-.
\eeq
Again we see that the expansion remains valid as long as $q^2$ scales
with the heavy masses (squared), and this limitation arises solely
from the coupling of the photon to the light quark.

\subsection{Spin Symmetry}
The NRQCD lagrangian contains separate fields for the charm quark and
antiquark.  The quark lagrangian, Eq.~(\ref{eq:NRQCDlag}) is
symmetric under spin-$SU(2)$ transformations. The antiquark
lagrangian is similarly invariant under a separate
spin-$SU(2)$. This case has a larger spin symmetry than the
case of decays to $D$-mesons. One can therefore write a trace formula
analogous to Eq.~(\ref{eq:werels}) without using chiral symmetry of
the light quarks.

We can represent the charmonium spin multiplet $(\eta_c,J/\psi)$ by
the $4\times4$ matrix
\beq
H^{(\psi)}_{v'}=
\left(\frac{1+\sla{v}'}{2}\right)
[\psi_\mu\gamma^\mu-\eta_c\gamma_5]
\left(\frac{1-\sla{v}'}{2}\right).
\eeq
The action of  spin-$SU(2)\times SU(2)$ on this is then
\beq
\label{eq:Hpsispin}
H^{(\psi)}_{v'}\to   S_c H^{(\psi)}_{v'} S^\dagger_{\bar c}
\eeq

Consider the matrix element $\vev{H^{(\psi)}_{v'}|\tilde{\cal
O}|H^{(b)}_v}$. It must be linear in the tensors $\Gamma_c\otimes
\Gamma_b$, $H^{(b)}_v$ and $\bar H^{(\psi)}_{v'}$. As before, acting with
$SU(2)_b$ we see that $\Gamma_b\to \Gamma_b S_b^\dagger$ and
$H_v^{(b)}\to S_b H_v^{(b)}$, so they enter the matrix element as
$\Gamma_bH_v^{(b)}$. Now, acting with the spin symmetries of NRQCD, we
have Eq.~(\ref{eq:Hpsispin}) and $\Gamma_c\to  S_c \Gamma_c
S^\dagger_{\bar c}$, so that they must enter the matrix element as 
${\rm Tr} (\bar H^{(\psi)}_{v'}\Gamma_c)$. Finally, rotations demand
that we sum over the two remaining indices,
\beq
\label{eq:werelscharmonium}
\vev{H^{(\psi)}_{v'}|\tilde{\cal O}|H^{(b)}_v} =\frac14\beta
{\rm Tr}(\bar H^{(\psi)}_{v'} \Gamma_c) {\rm Tr} (\Gamma_b H^{(b)}_v).
\eeq
Similarly, for the octet operator we find
\beq
\label{eq:werels8charmonium}
\vev{H^{(\psi)}_{v'}|\tilde{\cal O}_8|H^{(b)}_v} =\frac14\beta_8(w)
{\rm Tr}(\bar H^{(\psi)}_{v'} \Gamma_c) {\rm Tr} (\Gamma_b H^{(b)}_v).
\eeq
We have used the same symbols here for  operators and reduced matrix
elements as in Secs.~\ref{sec:method} and~\ref{sec:spinsym}, but they
should be understood as distinct.

\subsection{QCD Corrections}
\label{sec:QCD2}
Consider the operator expansion~(\ref{eq:OPEfullcharmonium}). Just as
in Sec.~\ref{sec:QCD1} we argue that matching between left and right
sides is most conveniently performed when the renormalization point
$\muz$ is chosen to be of the order of the scale of the heavy
quarks. For simplicity we assume that $m_c$ and $m_b$ are not too
different, but very big, so that we do not have to worry about large
logs of the ratio $m_c/m_b$. Then one may take, say,
$\muz\sim\sqrt{m_cm_b}$. The point is that the coefficients on the
left hand side of~(\ref{eq:OPEfullcharmonium}) explicitly depend on
$M_W/\muz$ and the operators implicitly depend on $m_{c,b}/\muz$.
If we choose to do the matching at a scale $\muz$ that differs much
from $m_{c,b}$ then there are implicit large corrections. Note that
the right hand side of~(\ref{eq:OPEfullcharmonium}) can only introduce
logs of low scales over $\muz$, but the same infrared logs are found
on the left side of the equation.

Once the coefficients $\tilde c$ and $\tilde c_8$
in~(\ref{eq:OPEfullcharmonium}) have been determined at $\muz$ we must
ask at what scale $\mu$ we should evaluate the matrix elements and how
to get there. The situation is more complicated than in the case of
$B^0\to D^0e^+e^-$ of Sec.~\ref{sec:QCD1} because now the matrix
element in the combined HQET/NRQCD effective theory has several
scales. In NRQCD the relevant distance scale is the inverse Bohr
radius $m_c\alpha_s(m_c)$ and the relevant temporal scale is the
Rydberg $m_c\alpha_s^2(m_c)$. In HQET the dynamical scale is
$\LQCD$. Of course $\LQCD$ also plays a dynamical role in NRQCD, but
it is usually taken to be irrelevant since one assumes $\LQCD\ll
m_c\alpha_s^2(m_c)\ll m_c\alpha_s(m_c)$. So we are faced with a
multiple scales problem. Setting $\mu$ equal to any one of these
scales leaves us with large logs of the ratios of $\mu$ to the other
two. It is not known how to use the renormalization group equation to
re-sum these logs.

Suppose that we set $\mu\sim m_c\alpha(m_c)$ or $\mu\sim
m_c\alpha^2(m_c)$. If we then use the renormalization group to sum
powers of $\alpha(m_c)\ln(m_c/\mu)$ we will be summing powers of
$\alpha(m_c)\ln\alpha(m_c)$. Notice that these logs vanish as
$m_c\to\infty$, since $\alpha(m_c)\sim 1/\ln(m_c/\LQCD)$. Contrast
this with the case $\mu\sim\LQCD$ (or, generally, setting $\mu$ equal
to any fixed scale as $m_c\to\infty$). Then
$\alpha(m_c)\ln(m_c/\mu)\sim1$ as $m_c\to\infty$. As a matter of
principle, in the large mass limit it is these latter logs that must
be summed (they are parametrically of leading order in the large mass
expansion). Therefore we re-sum the leading logs with a fixed low scale
$\mu=\mu_{\rm low}$ and choose, as before, $\mu_{\rm low}=1.0$~GeV in
our numerical computations.

In order to use dimensional regularization and keep track of different
orders in the non-relativistic expansion we adopt the $1/c$ counting
advocated in Ref.~\cite{bgir}. However, we use a covariant gauge for
our calculations. This is convenient because the Feynman diagrams
involve light and HQET quarks in addition to the NRQCD quarks. In
leading order in the $1/c$ expansion the quark lagrangian in
(\ref{eq:NRQCDlag}) is replaced by
\beq
\label{eq:NRQCDlagmod}
{\cal L}_{\rm NRQCD}\to \Psi^\dagger(iD_t -\frac{{\bf \nabla}^2}{2m_c})\Psi
\eeq
The only interactions are due to temporal gluon exchange. Since we
work in covariant gauge, this is not a pure Coulomb potential
gluon. It is easy to see that no diagram involving an NRQCD quark
gives a divergent contribution. The self-energy diagrams for the NRQCD
quarks have an infinite piece, which however is independent of the
momentum and therefore gives no contribution to wavefunction
renormalization.  Therefore the four quark operators scale as the
heavy-light currents. That is
\begin{equation}
\gamma=\frac{\alpha_s}{4\pi}
\left(\begin{array}{cc}
4 & 0\\
0 & 1 \\
\end{array}\right),
\end{equation}
is the anomalous dimension matrix in the renormalization group
equation for the operators,
\beq
\label{eq:rgeopscharmonium}
\mu\frac{d}{d\mu}
\left(
\begin{array}{c}
\tilde{\cal O} \\
\tilde{\cal O}_8 \\
\end{array}\right)
=\gamma \left(
\begin{array}{c}
\tilde{\cal O} \\
\tilde{\cal O}_8 \\
\end{array}\right).
\eeq
Then the coefficients must satisfy
\beq
\label{eq:rgecoefscharmonium}
\mu\frac{d}{d\mu}
\left(
\begin{array}{c}
\tilde{c} \\
\tilde{c}_8 \\
\end{array}\right)
=-\gamma^T \left(
\begin{array}{c}
\tilde{c} \\
\tilde{c}_8 \\
\end{array}\right),
\eeq
where, as above,  ``$T$'' denotes transpose of a matrix. 

The solution is trivial,
\begin{eqnarray}
\label{eqs:coefscharmonium}
\tilde c(\mu) & = & z^{a_I} \tilde c(\muz) \\
\label{eqs:coefscharmonium2}
\tilde c_8(\mu) & = & z^{\frac14a_I} \tilde c_8(\muz),
\end{eqnarray}
where $z$ is defined in Eq.~(\ref{eq:zdef}) and $a_I=2/b_0$ is the
well known anomalous scaling power for the heavy-light current in
HQET\cite{fggw}.

Contributions from higher orders in the $1/c$ expansion produce mixing
with higher dimension operators and are therefore excluded to the
order we are working. This is easy to see. To compensate for the
powers of $1/c$ one must have additional velocities in the
operators. But these come from powers of $\partial/m_c$. The leading
correction to the lagrangian is of order $1/c^{3/2}$. Since two
insertions are needed this gives a graph of order $1/c^3$. Since one
power of $c$ is needed to form the QCD fine-structure constant,
$\alpha_s=g_s^2/4\pi c$, the divergent part of the graph involves
$p^2/m_c^2c^2$. It is straightforward to verify this by direct calculation.

\subsection{Rates}

Defining 
\beq
h^{(\Psi)\mu}=
\langle \Psi| \int d^4x\;e^{iq\cdot x}
\; T(j^\mu_{\rm em}(x){\cal H}'_{\rm eff}(0)) |B_s\rangle,
\eeq
where $\Psi=\eta_c,J/\psi$, the decay rate for $B_s\to \Psi e^+e^-$ is
given in terms of $q^2$ and
$t\equiv(p_\Psi+p_{e^+})^2=(p_B-p_{e^-})^2$ by
\begin{equation}
\label{eq:doublediffratecharmonium}
\frac{d\Gamma}{dq^2dt}=\frac1{2^8\pi^3M_B^3}
\left|\frac{e^2}{q^2}\ell_\mu h^{(\Psi)\mu} \right|^2
\end{equation}
where $\ell^\mu=\bar u(p_{e^-})\gamma^\mu v(p_{e^+})$ is the leptons'
electromagnetic current. A sum over final state lepton helicities, and
polarizations in the $\Psi=J/\psi$ case, is implicit. 

We obtain
\begin{eqnarray}
h^{(\eta_c)\mu}=\frac\kappa3& &\Big[
\frac{m_bv^{\prime\mu}-2(wm_b-m_c)v^\mu}{(m_bv-2m_cv')^2}
\nonumber\\
& &\hspace{3cm}+\frac{m_bv^{\prime\mu}+2m_cv^\mu}{m_b^2-4m_c^2}\Big]
\end{eqnarray}
and 
\begin{eqnarray}
h^{(J/\psi)\mu}&=&\frac\kappa3\Big[
\frac{2m_bv\cdot\epsilon v^{\mu}-(m_b-2m_cw)\epsilon^\mu
-2m_cv\cdot\epsilon v^{\prime\mu}}{(m_bv-2m_cv')^2}\nonumber\\
&+&
\frac{2im_c\epsilon^{\mu\alpha\beta\gamma}\epsilon_\alpha v_\beta
v'_\gamma}{(m_bv-2m_cv')^2}
+\frac{8im_c\epsilon^{\mu\alpha\beta\gamma}\epsilon_\alpha v_\beta v'_\gamma}%
{m_b^2-2m_bm_cw}
\\
& &\hspace{-1cm}- \frac{m_b\epsilon^\mu+2m_c(v\cdot\epsilon
v^{\prime\mu}-w\epsilon^\mu)
+2im_c\epsilon^{\mu\alpha\beta\gamma}\epsilon_\alpha v_\beta v'_\gamma}%
{m_b^2-4m_c^2}
\Big].\nonumber
\end{eqnarray}
Here $\kappa=G_F/\sqrt2\,V^{\phantom{*}}_{cb}V_{cs}^*[\tilde
c\beta+\tilde c_8\beta_8]$. These expressions are our central results
for decays to charmonium, demonstrating that the decay rates for
$B_s\to \eta_c e^+e^-$ and $B_s\to J/\psi e^+e^-$ can be expressed in
terms of the local operator matrix elements $\beta$ and $\beta_8$.

We now compute the differential decay rate. We integrate the rate in
Eq.~(\ref{eq:doublediffratecharmonium}) over the variable $t$ and
obtain, for both $B_s\to \eta_c e^+e^-$  and $B_s\to J/\psi e^+e^-$,
\beq
\label{eq:rateBpsiee}
\frac{d\Gamma}{dq^2}=
\frac{\alpha^2 G_F^2}{288\pi M_B^3}|V_{cb} V_{cs}|^2 
(\tilde c\beta+\tilde c_8\beta_8)^2{\cal F}(\hq).
\eeq
Here ${\cal F}(\hq)$ is a dimensionless function of $\hq \equiv
\sqrt{q^2/m_b^2}$ and $\hm\equiv M_{J/\psi}/M_B$. For $B_s\to
J/\psi e^+e^-$ it is given by
\begin{eqnarray}
{\cal F} &=&
 \frac{4\sqrt{1-2 {\hq}^{2}-2 {\hm}^{2}+{\hq}^{4}
-2{\hm}^{2}{\hq}^{2}+{\hm}^{4}}}{3\hq^6{\hm}^{2} (1-\hm^2)^{2}
  (1+{\hq}^{2}-{\hm}^{2} )^{2}   }\nonumber\\
 & &(15 {\hm}^{10}-6 {\hm}^{12}+{\hm}^{2}+
15 {\hm}^{6}+{\hq}^{2}-6 {\hm}^{4}-20 {\hm}^{8}\nonumber\\
& &+{\hq}^{10}-2 {\hq}^{6}+6 {\hm}^{2}{\hq}^{2}+23 {\hm}^{
2}{\hq}^{4}+55 {\hm}^{2}{\hq}^{8}+{\hm}^{2}{\hq}^{12}\nonumber\\
& &-111 
{\hm}^{8}{\hq}^{2}+234 {\hm}^{6}{\hq}^{4}
+104 {\hm}^{6}{\hq}^{2}+92 {\hm}^{6}{\hq}^{6}\nonumber\\
& &-188 {\hm}^{8}
{\hq}^{4}+58 {\hm}^{10}{\hq}^{2}-46 {\hm}^{4}{\hq}^{2}+30
 {\hm}^{4}{\hq}^{6}\nonumber\\
& &-124 {\hm}^{4}{\hq}^{4}+{\hm}
^{14}-72 {\hm}^{8}{\hq}^{6}+55 {\hm}^{10}{\hq}^{4}-12 {
\hm}^{12}{\hq}^{2}\nonumber\\
& &+23 {\hm}^{6}{\hq}^{8}-78 {\hm
}^{4}{\hq}^{8}+4 {\hm}^{4}{\hq}^{10}-6 {\hm}^{2}{\hq}^{10}
-48 {\hm}^{2}{\hq}^{6} ) \nonumber\\
\end{eqnarray}
while for $B_s\to \eta_c e^+e^-$
\beq
{\cal F} =
\frac{4(1-2 \hq^2-2 {\hm}^{2}+{\hq}^{4}
-2 {\hm}^{2}{\hq}^{2}+{\hm}^{4})^{\frac32}}
{3\hq^4 \hm^2 (1-\hm^2)^2}.
\eeq

For a numerical estimate we need to calculate the matrix elements
$\beta$ and $\beta_8$. Again we use vacuum saturation. However, now
this approximation is supported by NRQCD. It is argued in
Ref.~\cite{bbl} that soft gluon exchange with the quarkonium is
suppressed by powers of the relative velocity $u=\alpha_s(m_c)$, and
that the matrix element of the octet operator is similarly
suppressed. Therefore we take
\begin{eqnarray}
\label{eqs:melemscharmonium}
\beta(w) & = &  z^{-a_I}f_Bf_{\eta_c}\sqrt{M_BM_{\eta_c}}\\
\label{eqs:melemscharmonium2}
\beta_8(w) & = &0.
\end{eqnarray}
Note that because vacuum saturation here is valid at least as a
leading approximation in a velocity expansion, the combination of
coefficients in
(\ref{eqs:coefscharmonium})--(\ref{eqs:coefscharmonium}) and matrix
elements in
(\ref{eqs:melemscharmonium})--(\ref{eqs:melemscharmonium2}) is
automatically independent of the renormalization point $\mu$.  Spin
symmetry gives $f_{\eta_c}=f_{J/\psi}$. We use the measured value from
the leptonic width in the tree level rate equation,
\beq
\Gamma(J/\psi\to e^+e^-)=4\pi\alpha^2\frac{f_{J/\psi}^2}{M_{J/\psi}},
\eeq
and obtain $f_{J/\psi}=0.16$~GeV.

Integrating over $q^2\ge1.0$~GeV we have partial branching fractions
\begin{eqnarray}
\label{eq:btojpsirate}
{\rm Br}(B_s\to J/\psi e^+e^-)|_{q^2>1~{\rm GeV}}
&=&2.2\times10^{-10}\\
{\rm Br}(B_s\to \eta_c e^+e^-)|_{q^2>1~{\rm GeV}}
&=&3.4\times10^{-11}
\end{eqnarray}
where we have used $|V_{cb}V_{cs}|=0.04$,  and
$f_B=170$~MeV. Again, we remind the reader that the portion
of phase space $q^2\ge 1.0~{\rm GeV}$ is a small fraction
of the total rate since the pole at $q^2=0$ dramatically amplifies
the rate for small $q^2$. The rates for $B^0\to J/\psi  e^+e^-$ and
$B^0\to\eta_c e^+e^-$ can be obtained to good approximation by
replacing $|V_{cb}V_{cs}|$ by $|V_{cb}V_{cd}|$, reducing the rates by
$(0.22)^2\approx0.05$. The rate (\ref{eq:btojpsirate}) may seem  too small
to be detectable even in the next generation of hadronic
colliders. However it must be kept in mind that the signature involves
four leptons with large invariant masses (one being the $J/\psi$).

\section{Conclusions}
\label{sec:conclusions}
We have successfully shown how to implement the OPE advertised in 
Ref.~\cite{evans-99-2} to the processes $\bar B_{d,s}\to J/\psi e^+e^-$,
 $\bar B_{d,s}\to \eta_c e^+e^-$, $\bar B_{d,s}\to D^{*0} e^+e^-$ and 
$\bar B_{d,s}\to D^{0} e^+e^-$. By the use of the OPE the long distance 
(first order weak and first order electromagnetic) interaction  is replaced 
by a sum of local operators. The application of the OPE is restricted 
to a limited kinematic region.

In the processes $\bar B_{d,s}\to J/\psi e^+e^-$ and 
$\bar B_{d,s}\to \eta_c e^+e^-$ our method leads naturally to an 
NRQCD expansion for the $J/\psi$ and $\eta_c$. This illustrates that 
the methods of Ref.~\cite{evans-99-2} are applicable to a wider class 
of processes.

Furthermore we found that the number of independent matrix elements of 
the local operators is severely restricted due to a combined use of 
heavy-spin, rotational and chiral symmetry. The independent matrix 
elements could be determined, say, in lattice simulations. 
Our paper shows that 
the processes considered can be studied in a systematic fashion independent 
of any model assumptions in 
the kinematic regime of $q^2$ scaling like $m_{c,b}^2$.

Using a crude estimation of the  matrix elements, we found
the rates of all the processes considered to be small.  We expect some
of them, in particular $\bar B_{s}\to D^{*0} e^+e^-$, should be
accessible at planned experiments at hadron colliders, like BTeV or
LHC-B.

\bigskip

{\it Acknowledgments} 
We thank Mark Wise for useful discussions and conversations.
This work is supported by the Department of
Energy under contract No.\ DOE-FG03-97ER40546.

\tighten

\end{document}